\documentstyle[12pt]{article} 
\def\beq{\begin{equation}} 
\def\eeq{\end{equation}}
\def\bea{\begin{eqnarray}} 
\def\eea{\end{eqnarray}}
\def\bq{\begin{quote}} 
\def\eq{\end{quote}}
\def\gappeq{\mathrel{\rlap {\raise.5ex\hbox{$>$}}
{\lower.5ex\hbox{$\sim$}}}}

\def\lappeq{\mathrel{\rlap{\raise.5ex\hbox{$<$}}
{\lower.5ex\hbox{$\sim$}}}}
\begin{document} 
\begin{titlepage}
\begin{flushright} 
{CERN-TH.98-226\\
DFPD-98/TH-34}
\end{flushright} 
\vspace*{5mm}
\begin{center} 
{\bf Neutrino Mass Textures
from 
Oscillations with Maximal Mixing} 
\end{center}
\vskip 0.8 cm
\begin{center}
{\bf Guido Altarelli} 
\\ 
\vspace{0.3cm} Theoretical Physics Division, CERN 
\\
CH - 1211 Geneva 23 
\\ 
and
\\
Universit\`a di Roma Tre, Rome, Italy 
\\
\vskip 0.5 cm 
{\bf Ferruccio Feruglio} 
\\ 
\vspace{0.3cm}  
Universit\`a di Padova, Padova, Italy
\\
and
\\
I.N.F.N., Sezione di Padova
\end{center} 
\vspace*{1.5cm}   
\begin{abstract}
\noindent 
We study the implications of neutrino oscillations with maximal mixing for
the neutrino
Dirac and Majorana matrices in the see-saw mechanism for three non
degenerate neutrino masses. We find the form of the
Dirac matrix for a structure-less Majorana matrix and, conversely, the
structure of the Majorana matrix if the
Dirac matrix is according to our naive intuition. We give some examples of
Majorana matrices that, in a $3\times 3$ context,
lead to maximal mixing without too much fine tuning and cross talk
with the Dirac input.
\end{abstract} 
\vspace*{5mm} \noindent

\vspace*{2.5cm}  \noindent  

\noindent

\begin{flushleft} CERN-TH.98-226 \\ DFPD-98/TH-34 \\July 1998 \end{flushleft} 
\end{titlepage}

\setcounter{page}{1} \pagestyle{plain}



\section{}

Recent data from Superkamiokande \cite{SK}(and also MACRO \cite{MA}) have
provided a more solid
experimental basis for neutrino oscillations as an explanation of the
atmospheric neutrino anomaly.
In addition, the solar neutrino deficit is also probably an indication of a
different sort of neutrino
oscillations. Results from the laboratory experiment by the LNSD
collaboration \cite{LNSD} can also be
considered as a possible indication of yet another type of neutrino
oscillation. But the preliminary
data from Karmen \cite{KA} have failed to reproduce this evidence. The
case of LNSD oscillations is far
from closed but one can tentatively assume, pending the results of
continuing experiments, that the
signal will not persist. Then solar and atmospheric neutrino oscillations
can possibly be explained in
terms of the three known flavours of neutrinos without invoking extra
sterile species. Neutrino
oscillations for atmospheric neutrinos require
$\nu_{\mu}\rightarrow\nu_{\tau}$ with $\Delta m^2_{atm}\sim
2\cdot 10^{-3}~eV^2$ and a nearly maximal mixing
angle
$\sin^2{2\theta_{atm}}\geq 0.8$. Furthermore the last results from
Superkamiokande allow \cite{Bahcall} for a
solution of the solar neutrino deficit in terms of $\nu_e$ disappearance
vacuum oscillations (as opposed to
MSW \cite{MSW} oscillations within the sun) with $\Delta m^2_{sol}\sim
~10^{-10}~eV^2$ and again nearly maximal mixing
angles. Among the large and small angle MSW solutions the small angle one
(with \cite{Bahcall} $\Delta m^2_{sol}\sim 0.5\cdot 10^{-5}~eV^2$ and
$\sin^2{2\theta_{sol}}\sim
5.5\cdot 10^{-3}$) is more likely  at the
moment  than the large angle MSW solution. Of course experimental
uncertainties are still large and the
numbers given here are merely indicative. Thus atmospheric neutrinos, if a
genuine signal of oscillations, certainly
require nearly maximal mixing and possibly also solar neutrinos may arise
from nearly maximal mixing. Large mixings
are very interesting because a first guess was in favour of small mixings
in the neutrino sector in analogy to what
is observed for quarks. If confirmed, single or double maximal mixings can
provide an important hint on the
mechanisms that generate neutrino masses.

In the present note we start from the simplest (at least to our taste)
scenario with only three flavours of
neutrinos that receive masses from the see-saw mechanism, with a given
hierarchical structure for the
neutrino mass eigenvalues $m_i$ (e.g. $m_3\gg m_2\gg m_1$ or $m_3\gg m_2\sim
m_1$). We consider the simplest version of
the see-saw mechanism with one Dirac, $m_D$, and one Majorana, $M$, mass
matrix, related to the neutrino mass matrix
$m_{\nu}$, in the basis where the charged lepton mass matrix is diagonal,
by
\beq
m_{\nu}=m^T_D M^{-1}m_D
\label{1}
\eeq
As well known this is not the most general
see-saw mechanism because we are not including the left-left Majorana mass
block. This assumption is done here for
reasons of simplicity and because it represents the most constrained
situation: allowing this extra matrix would
leave more freedom. We study the implications of maximal mixing (either
single or double) on the Dirac and
Majorana matrix structure. One may
imagine that maximal mixing cannot be accommodated in a natural way in
presence of a large spread of the neutrino
masses and that maximal mixing goes in the direction of degenerate masses.
We want to investigate to what extent
this statement is really justified. We analyse in detail the most
interesting limiting cases. In one (sect.2) all
the structure arises from the Dirac matrix $m_D$ while $M$ is the simplest. We
find that the resulting Dirac matrix has
a well defined structure. Once this structure is realized then maximal
mixing is obtained without fine tuning. Or,
alternatively (sect.3), the Dirac matrix is taken according to the
intuition that it should be similar to the up
quark mass matrix, (for example, with small mixings among mass splitted
states, possibly with a remnant of the
top, charm, up mass pattern) and the large mixing(s) are induced by $M$.
For non degenerate masses what is obtained
is a texture of the Majorana matrix
$M_{ij}$ with increasing absolute values when i+j increases, which is more
pronounced for double versus single
maximal mixing. However in this case the realization of the specific
texture is not sufficient and extra fine
tuning is needed. More interesting solutions are found if we allow near
degeneracy for $m_1$ and $m_2$.

Our main framework is the most obvious case where the observed relation
$\Delta m^2_{atm}\gg\Delta m^2_{sol}$ is a
reflection of the mass hierarchy $m_3\gg m_{2,1}$, with no prejudice on the
$m_1$, $m_2$ relation. Maximal
atmospheric neutrino mixing and the requirement that the electron neutrino
does not participate in the atmospheric
oscillations, as indicated by the Superkamiokande \cite{SK} and Chooz
\cite{Chooz} data, lead directly to the
following structure of the
$U_{fi}$ (f=e,$\mu$,$\tau$, i=1,2,3) real orthogonal  mixing matrix, apart
from sign convention redefinitions (here
we are not interested in CP violation effects, which are small in our
context, being exactly zero for $U_{e3}=0$):
\beq
U_{fi}= 
\left[
\matrix{
c&-s&0 \cr
s/\sqrt{2}&c/\sqrt{2}&-1/\sqrt{2}\cr
s/\sqrt{2}&c/\sqrt{2}&+1/\sqrt{2}     } 
\right]
\label{2}
\eeq
This result is obtained by a simple generalization of the analysis of
ref.\cite{bar} to the case of arbitrary
solar mixing angle
\footnote{An analogous parametrization has been discussed in ref. \cite{Bal}.}
($s\equiv\sin{\theta_{sun}}$,
$c\equiv\cos{\theta_{sun}}$): $c=s=1/\sqrt{2}$ for maximal solar
mixing, while $\sin^2{2\theta_{sun}}\sim 4s^2\sim 5.5\cdot 10^{-3}$ for the
small angle MSW solution. The vanishing of
$U_{e3}$ guarantees that 
$\nu_e$ does not participate in the atmospheric oscillations and the
relation $|U_{\mu3}|=|U_{\tau3}|=1/\sqrt{2}$
implies maximal mixing for atmospheric neutrinos. The non diagonal
oscillation probabilities are
\bea
P(\nu_e\leftrightarrow\nu_{\mu})=
P(\nu_e\leftrightarrow\nu_{\tau})&=&
2c^2s^2\sin^2{\Delta_{sun}}\nonumber\\
P(\nu_{\mu}\leftrightarrow\nu_{\tau})&=&
\sin^2{\Delta_{atm}}-c^2s^2\sin^2{\Delta_{sun}}
\label{3}
\eea
Note that we are assuming only two frequencies, given by
$\Delta_{sun}\propto m^2_2-m^2_1$ and
$\Delta_{atm}\propto m^2_3-m^2_{1,2}$. A more general analysis can be
found in ref.\cite{barb}. The neutrino mass matrix
is given by
$Um_{diag}U^T$ with
$m_{diag}=Diag[m_1,m_2,m_3]$. In the following we will always give
$m_{\nu}$ in units of $m_3/2$, $m_D$ in units of
$m_0/\sqrt{2}$, and $M$ in units of some large mass $\bar M$ related to
$m_3$ and $m_0$ by $m_3=m^2_0/\bar M$.  With
this convention, for generic $s$ one finds
\beq
m_{\nu}= 
\left[
\matrix{2\epsilon&\delta&\delta\cr
\delta&1+\epsilon_2&-1+\epsilon_2\cr
\delta&-1+\epsilon_2&1+\epsilon_2} 
\right]
\label{4}    
\eeq
with 
\beq
\epsilon=(\mu_1 c^2+\mu_2 s^2),~~~\delta=\sqrt{2}(\mu_1-\mu_2) c
s,~~~\epsilon_2=(\mu_1 s^2+\mu_2 c^2)
\label{5}    
\eeq
where $\mu_i\equiv m_i/m_3$. We see that the existence of one maximal
mixing and $U_{e3}=0$ are the most important
input that leads to the matrix form in eq. (\ref{4}). The value of the
solar neutrino mixing angle can be left
free. 

\section{}

We consider first the case of the simplest Majorana matrix $M$, namely a
multiple of the identity: $M\equiv~1$. 
In the non degenerate case $m_3\gg m_2,m_1$ we find that eq. (\ref{1})
is satisfied by the following texture:
\beq
m_D=\left[
\matrix{
\lambda&\lambda&\lambda\cr
\lambda&\sim c&\sim-c\cr
\lambda&\sim-t&\sim t}
\right]~~~,~~~~~~~~~~~c^2+t^2=1
\label{6}    
\eeq
where $\lambda$ is a small parameter related to $\mu_1,\mu_2$ and $s$.
By $\sim c(t)$ we mean $c(t)+O(\lambda^n)$ and by $\lambda$ we mean 
$O(\lambda^n)$
with $n\ge 1$
\footnote{As a particular case, if we take $\sqrt{2} m_D$
equal to $m_\nu$ in eqs. (\ref{4},\ref{5}) with all
$\mu_i$ replaced by $\sqrt{\mu_i}$, then eq. (\ref{1}) is
identically satisfied.}.
We have obtained this texture starting from a Majorana matrix 
which is a multiple of the
identity. But a particularly remarkable feature
of this texture is that it satisfies
eq.(\ref{1}) for whatever symmetric non singular
Majorana matrix with all matrix elements of order 1. Precisely, if we
denote by $m^0_D$ the matrix obtained by
setting $\lambda=0$ in eq.(\ref{6}), then, up to a multiplicative factor,
$m^0_D$ provides an exact solution of eq. (\ref{1}) 
for any symmetric non singular Majorana matrix. This
feature arises because the diagonal form of $m^0_D$ has only the entry 33
different from zero and hence is
proportional to a projector.
For $M=1$ an example of double maximal mixing is given by
the matrix
\beq
m_D=1/\sqrt{2}\left[
\matrix{
2\lambda&\lambda&\lambda\cr
\lambda&1+\lambda&-1+\lambda\cr
\lambda&-1+\lambda&1+\lambda} 
\right] 
\label{6a}    
\eeq
with $\lambda^2\sim10^{-4}-10^{-5}$. Note that an attractive feature of
double maximal mixing is that for 
$m_3\gg m_{1,2}$, $\mu_2$  
comes remarkably close to $m^2_c/m^2_t$. Also note that in general, if we
set $m_3\sim m^2_t/(\bar M)$, then $\bar M$
is close to the unification mass $m_{GUT}$.  Thus presumably
$\lambda^2$ could be related to
$m^2_c/m^2_t$. Similarly, an example of single maximal mixing is given by
the matrix
\beq
m_D=1/\sqrt{2}\left[
\matrix{
0&\lambda^2&\lambda^2\cr
\lambda^2&1+\lambda&-1+\lambda\cr
\lambda^2&-1+\lambda&1+\lambda} 
\right] 
\label{6b}    
\eeq
with $\lambda^2\sim10^{-3}$.
The pattern of eq. (\ref{6}) can be also specialized to the case
$t\gg c$. For instance, for $M=1$ a double maximal 
mixing solution is offered by the matrix
\beq
m_D=\left[
\matrix{
\lambda^2&\lambda&0\cr
\lambda&\lambda&\lambda\cr
0&-1&1} 
\right] 
\label{6f}    
\eeq
with $\lambda^2\sim 10^{-4}$. A further example of single maximal mixing
is provided by the matrix:
\beq
m_D=\left[
\matrix{
\lambda^2&\lambda^2&0\cr
\lambda^2&\lambda&\lambda\cr
0&-1&1} 
\right] 
\label{6g}    
\eeq
with $\lambda^2\sim 10^{-2}$. In all the above examples, eqs. 
(\ref{6a}-\ref{6g}), we have complete non degeneracy 
$m_3^2\gg m_2^2\gg m_1^2$.

We thus conclude that in the non degenerate case to explain atmospheric
neutrinos in terms of
$\nu_{\mu}\leftrightarrow \nu_{\tau}$ oscillations with maximal mixing,
for a non special Majorana matrix, one
needs a Dirac mass matrix with the texture in eq.(\ref{6}). This texture
is independent of the amount of mixing
for solar neutrinos. 
Thus if a natural way
of generating the texture in
eq. ({\ref{6}) is found the problem is solved. Examples of strategies to
approach this problem are given in
refs.\cite{babu,elwood}. But if one insists that large mixings are
excluded in the Dirac sector then one can
try to introduce a special texture in the Majorana sector. The required
Majorana texture is discussed in the next
section.

\section{}
 
We want to determine the Majorana mass matrix for the simplest and most
intuitive configuration for the Dirac 
matrix. So we take $m_D$ close to diagonal and given by
\beq
m_D=\left[
\matrix{
u&0&0\cr
0&c&d\cr
0&f&t} 
\right] 
\label{6c}    
\eeq
where the zeroes represent smaller entries. For the time being the
magnitudes of $u$, $c$ and $t$ are not specified but the notation 
is suggestive
of up, charm and top quarks. For $f=d$ the general
form of $M$ that reproduces $m_{\nu}$ in eqs.(\ref{4}-\ref{5}) via
eq.(\ref{1}) is given by
\beq
M=1/D\left[
\matrix{
\epsilon_2u^2&-\frac{\delta}{2}u(c+d)&-\frac{\delta}{2}u(t+d)\cr
-\frac{\delta}{2}u(c+d)&\frac{D}{4}(c-d)^2+\frac{\epsilon}{2}(c+d)^2&-
\frac{D}{4}(c-d)(t-d)+\frac{\epsilon}{2}(c+d)(t+d)\cr
-\frac{\delta}{2}u(t+d)&-\frac{D}{4}(c-d)(t-d)+\frac{\epsilon}{2}(c+d)(t+d)&
\frac{D}{4}(t-d)^2+\frac{\epsilon}{2}(t+d)^2} 
\right]
\label{17}    
\eeq
where
\beq
D= \frac{Det[m_{\nu}]}{4}= 2\epsilon\epsilon_2-\delta^2=2\mu_1\mu_2
\label{18}    
\eeq
Note that $Det[M]=(Det[m_D])^2/Det[m_{\nu}]=[u(ct-d^2)]^2/(8\mu_1\mu_2)$.
If the normalizing factor $\bar M$ is
appropriately chosen, without loss of generality, we can take $Det[M]$ of
order 1, and then eq.(\ref{18}) is an
important constraint on the relative magnitudes of $u$, $c$, $d(=f)$ and $t$ 
vs $\mu_1$ and $\mu_2$. 

We see immediately from eq.(\ref{17}) that, for $t\gg c,u$ and
$\epsilon\sim\epsilon_2\sim\delta$ small, $D$ is in general negligible and
the matrix elements $M_{ij}$
tend to increase when $i+j$ increases. For example, consider the case of
double maximal mixing (vacuum solution
for solar neutrinos). The gap between $\Delta m^2_{atm}$ and $\Delta
m^2_{sol}$ is particularly large in this
case and suggests a pronounced hierarchy of neutrino masses. If we take
$\mu_2\sim\lambda^2$, $\mu_1\sim\lambda^4$ and
correspondingly
$u\sim\lambda^2$, $c,d(=f)\sim\lambda$ and $t\sim1$, then we have
$\epsilon\sim\epsilon_2\sim\delta\sim\lambda^2$ and
$D\sim\lambda^6$. The resulting texture for $M$ is given by
\beq
M=\left[
\matrix{
1&\frac{1}{\lambda}&\frac{1}{\lambda^2}\cr
\frac{1}{\lambda}&\frac{1}{\lambda^2}&\frac{1}{\lambda^3}\cr
\frac{1}{\lambda^2}&\frac{1}{\lambda^3}&\frac{1}{\lambda^4}} 
\right]
\label{19}    
\eeq
If we instead take $\mu_2\sim\mu_1\sim\lambda^2$ and $u\sim c\sim\lambda$
the entries in $M$ are less singular, i.e. the pattern is attenuated.

In the case of the small MSW solar neutrino solution, $\Delta m^2_{sol}$
is larger by about 5 orders of
magnitude, thus for $m_3\gg m_2\gg m_1$, $\mu_2$ is larger by about 2.5 orders
of magnitude  and $\lambda$ is a factor
of $\sim20$ larger. Also $\delta$ is suppressed by the small mixing angle
$s$, and numerically $s\sim\lambda^2$. In
this case $\epsilon\sim\mu^2_1$ could be smaller than
$\epsilon_2\sim\mu^2_2$. These features also result in an
attenuation of the texture. If we take $\mu_2\sim\lambda^2$,
$\mu_1\sim\lambda^4$ and correspondingly
$u\sim\lambda^2$, $c,d(=f)\sim\lambda$ and $t\sim1$, then we can now have
$\epsilon\sim\lambda^4$, $\epsilon_2\sim\lambda^2$, $\delta\sim\lambda^4$
and
$D\sim\lambda^6$. The resulting texture for $M$ is given by
\beq
M=\left[
\matrix{
1&\lambda&1\cr
\lambda&1&\frac{1}{\lambda}\cr
1&\frac{1}{\lambda}&\frac{1}{\lambda^2}} 
\right] 
\label{20}    
\eeq
If we instead take $\mu_2\sim\mu_1 \sim \lambda^2$ and $u\sim
c\sim\lambda$ with
$\epsilon\sim\epsilon_2\sim\lambda^2$ and $\delta\sim\lambda^4$ one finds
that the texture is further attenuated.

The essential difference with the Dirac case is that, in the Majorana
case, while the above type of textures are
what comes out in absence of special relations among the coefficients,
they are not sufficient. The ratios 
of the different entries have to be tuned in the way shown by the
general expression in eq.(\ref{17}), not only in order to obtain a
determinant of order 1, but also to ensure the
additional delicate cancellations needed to lead to the correct mixings. In
other words, the nearly diagonal Dirac
matrix and the Majorana matrix must together conspire in order to produce
the result. Some crosstalk between the
two matrices can be induced by the diagonalization of the charged lepton
mass matrix (that changes $M^{-1}\rightarrow
VM^{-1}V^T$). But in the logic of small Dirac mixing this interconnection
should be limited to small terms. 

The only
possibility for a relatively disconnected first approximation is to go to
some limiting form for the Majorana matrix,
obtained by playing with the parameters $\epsilon$, $\epsilon_2$ and
$\delta$. We find that this is only possible if
$\mu_1$ and $\mu_2$ are nearly degenerate, because then $\delta$ can be of
different order than the $\epsilon$'s. We
want a texture for $M$ of O(1) and O(0) entries, as much as possible
independent of the $m_D$ matrix elements, that in
conjunction with $m_D$ given in eq.(\ref{6c}) produce large mixing(s).
Solutions for the Majorana matrix can be obtained
by discussing the limiting case where the small quantities
$\epsilon,\epsilon_2,\delta$ vanish.
Then, eq. (\ref{1}) represents a set of constraints on the most
general symmetric and invertible matrix $M$.

For instance, at first we demand small non diagonal terms in the 
Dirac matrix. By taking in eq. (\ref{6c}) $c=t=1$ and by setting,
in first approximation, $u=d=f=0$ (i.e. $m_D=Diag[0,1,1]$),
we obtain the result:
\beq
M=\left[
\matrix{
0&C&C\cr
C&2A-B+1&A\cr
C&A&B} 
\right] 
\label{21}   
\eeq
Here A,B and C ($\ne0$) are independent parameters. 
At this point we can turn on small entries in the Dirac matrix.
For instance, if we take $A,B,C\sim O(1)$, $u\sim\lambda$, 
$d\sim f\sim k\lambda$,
by small readjustments and by taking $\lambda$ close to $m_c/m_t$ 
we can find an acceptable doubly mixed solution with $m_3$ large
and $\mu_1$ and $\mu_2$ nearly degenerate. 
This recipe is somewhat similar to
the one proposed in a $2\times 2$ context 
in ref. \cite{alla}. 

Alternatively we can try to restore the hierarchy
$t\gg c$, by taking $m_D$ symmetric but non diagonal. In fact, for 
$d=f=t=1$ and $u=c=0$, we obtain the result
\beq
M=\left[
\matrix{
0&C&2C\cr
C&B&A\cr
2C&A&4A-4B+1} 
\right] 
\label{22}    
\eeq
where $A,B$ and $C(\ne 0)$ are independent quantities.
Again, for $c\sim\lambda$ and $u$ small, 
this corresponds to double maximal mixing. 

An additional class of solutions with the Dirac matrix
quasi diagonal and hierarchical can be obtained starting from
a non symmetric form of $m_D$: $d=-c$ and $f$ generic in eq. (\ref{6c}).
For $u=0$, we find that
\beq
M=\left[
\matrix{
0&0&C\cr
0&D&A\cr
C&A&B} 
\right] 
\label{24}    
\eeq
solves eq. (\ref{1}) up to the overall factor $c^2/D$
that can be absorbed in the definitions of $m_0$ and ${\bar M}$.
By considering a small non-vanishing $u$, $m_\nu$ acquires 12 and 13
entries of order $u/c^2$ compared to the entries in the 2-3 block.
A remarkable feature of this solution is that
all what is needed  in order to reproduce the desired pattern
is to set to zero three elements of $M$, as displayed
in eq. (\ref{24}). The remaining quantities $A$, $B$,
$C(\ne 0)$ and $D(\ne 0)$ are completely independent from the Dirac data.
By taking $A$, $B$, $C$, $D$, $t$ of $O(1)$, 
$f\sim\sqrt{\lambda}$, $c(=-d)\sim\lambda$ and $u\sim\lambda^3$
with $\lambda\sim 10^{-2}$ we find an acceptable solution
with double maximal mixing. 
A particular case $A=B=0$ of this pattern has been considered in ref.
\cite{sum}. For the last two cases, eqs.
(\ref{22}) and (\ref{24}), the ratio
$m_2/m_3$ is in general not related to $c/t$ although
numerically it can be tuned to be so. 

It is interesting that we find that
double maximal mixing is simpler to obtain
in this framework (because both $\epsilon$ and $\epsilon_2$ are smaller
than $\delta$ and their difference does not
appear). On the other hand we do not accept to fine tune $s$ in eqs.(\ref{5})
in terms of $\mu_1$ and $\mu_2$ in order to
make one $\epsilon$ of different order than the other.
We also observe that, by appropriately choosing the parameters 
in eqs. (\ref{21}), (\ref{22}) and (\ref{24}), one can obtain special 
Majorana textures with up to six vanishing entries.

\section{}

In conclusion, we have studied the implications of neutrino 
oscillations with maximal mixing for the neutrino
Dirac and Majorana matrices in the see-saw mechanism. Maximal mixing can
be taken as 
an indication of nearly degenerate neutrino masses, because in
perturbation theory mixing 
is small among widely non degenerate states. For the widely spread quark
mass eigenstates 
the mixings are indeed small. In fact nearly degenerate neutrino masses
have been widely 
considered recently. One advantage of full near degeneracy is that the
three neutrinos can still
provide a component of hot dark matter if the common mass is around
$1~eV$. We think that it is not really clear
at the moment that a hot dark matter component is really needed 
\cite{kra}. So we
have studied the non degenerate case $m_3^2\gg m_{1,2}^2$ and we
have determined the conditions for maximal mixing to be imposed to the
Dirac and the Majorana matrices of the see-saw
mechanism. For a non special Majorana matrix the Dirac matrix structure is
simple and well defined. Opposite j2 and j3 $(j=2,3)$ matrix
elements are required in the $\nu_{\mu}$ - $\nu_{\tau}$
mass entries and small matrix elements elsewhere. But once this
requirement is met no additional fine tuning is
needed. There are examples of strategies that may lead to this pattern in
the context of grand unified theories \cite{elwood}. So
we do not consider that this possibility is to be discarded. On the
contrary even if the Dirac matrix does not involve
large mixings, the latter can still be obtained through the Majorana
matrix. We have exhibited some examples of
Majorana textures that lead to double maximal mixing without too much fine
tuning and cross talk with the Dirac input.
Further work is clearly needed to generate in a natural way the proposed
textures from some reasonable dynamical
context.


\begin{thebibliography}{99}


\bibitem{SK}
Y. Fukuda et al., hep-ex/9805006, hep-ex/9805021 and hep-ph/9807003.
\bibitem{MA}
M. Ambrosio et al., hep-ex/9807005.
\bibitem{LNSD}
C. Athanassopoulos et al., Phys. Rev. Lett. 77 (1996) 3082,
nucl-ex/9706006 and nucl-ex/9709006.
\bibitem{KA}
B. Armbruster et al., Phys. Rev. C57 (1998) 3414 and G. Drexlin, talk at 
Wein'98.
\bibitem{MSW}
L. Wolfenstein, Phys. Rev. D17 (1978) 2369;\\
S.P. Mikheyev and A. Yu Smirnov, Sov. J. Nucl. Phys. 42 (1986) 913.
\bibitem{Bahcall}
G.L. Fogli, E. Lisi and D. Montanino, hep-ph/9709473 and hep-ph/9803309;\\
B. Fa\"{\i}d, G.L. Fogli, E. Lisi and D. Montanino, hep-ph/9805293;\\
J.N. Bahcall, P.I. Krastev and A. Yu Smirnov, hep-ph/9807216.
\bibitem{Chooz}
M. Apollonio et al., Phys. Lett. B 420 (1998) 397.
\bibitem{bar}
V. Barger, S. Pakvasa, T.J. Weiler and K. Whisnant, hep-ph/980638;
V. Barger, T.J. Weiler and K. Whisnant, hep-ph/9807319.
\bibitem{Bal}
A.J. Baltz, A.S. Goldhaber and M. Goldhaber, hep-ph/9806540.
\bibitem{barb}
R. Barbieri, L.J. Hall, D. Smith, A. Strumia and N. Weiner, hep/ph 9807235.
\bibitem{babu}
K.S. Babu and S.M. Barr, Phys. Lett. B 381 (1996) 202 (hep-ph/9511446).
\bibitem{elwood}
G. Costa and E. Lunghi, Nuovo Cim. A110 (1997) 549;\\
B. Brahmachari and R.N. Mohapatra, Phys. Rev. D58 (1998) (hep-ph/9710371);\\
S.F. King, hep-ph/9806440;\\
J.K. Elwood, N. Irges and P. Ramond, hep-ph/9807228;\\
Y. Nomura and T. Yanagida, hep-ph/9807325.
\bibitem{alla}
B.C. Allanach, hep-ph/9806294.
\bibitem{sum}
M. Jezabek and Y. Sumino, hep-ph/9807310.
\bibitem{kra}
See for example L. K. Krauss, hep-ph/9807376.
\end{thebibliography}
\end{document}